\documentclass[12pt,a4paper]{article}
\usepackage{amssymb}

\begin{document}

\newcommand{\be}{\begin{equation}}
\newcommand{\ee}{\end{equation}}
\newcommand{\beann}{\begin{eqnarray*}}
\newcommand{\eeann}{\end{eqnarray*}}
\newcommand{\bea}{\begin{eqnarray}}
\newcommand{\eea}{\end{eqnarray}}
\newcommand{\lb}{\label}
\newcommand{\bdm}{\begin{displaymath}}
\newcommand{\edm}{\end{displaymath}}

\begin{titlepage}

\noindent

\vspace*{1cm}
\begin{center}
{\large\bf QUANTUM COSMOLOGY 
AND THE ARROW OF TIME}\footnote{Contribution
to the Proceedings of the conference DICE2004, Piombino, Italy, 
September 2004; to appear in {\em Brazilian Journal of Physics}.} 

\vskip 1cm

{\bf Claus Kiefer} 
\vskip 0.4cm
Institut f\"ur Theoretische Physik,\\ Universit\"{a}t zu K\"oln, \\
Z\"ulpicher Str.~77,
50937 K\"oln, Germany.\\
\vspace{1cm}

\begin{abstract}
Although most fundamental laws are invariant under time reversal,
experience exhibits the presence of irreversible phenomena -- the 
arrows of time. Their origin lies in cosmology, and I argue that only quantum
cosmology can provide the appropriate formal framework. 
After briefly reviewing the formalism, I discuss how a simple and
natural boundary condition can lead to the observed arrows of time.
This yields at the same time interesting consequences for black holes.
\end{abstract}
\end{center}

\end{titlepage}

%%%%%%%%%%%%%%%%%%%%%%%%%%%%%%%%%%%%%%%%%%%%%%%%
\section{Introduction}

Recent experiments support the idea that quantum theory
is universally valid. No breakdown of the superposition principle
has been detected, and the disappearance of interference term
can be understood in a quantitative way by the process of decoherence
\cite{deco}: Entanglement with environmental degrees of freedom
produces locally classical behaviour. The local system is then
consistently described only by a master equation, not a
unitary Schr\"odinger equation.
Apart from microscopic and
some mesoscopic systems, it is usually not possible to isolate a
system from its environment. Following this chain, the environment
is coupled to its environment, and so on, leading ultimately to the
whole Universe as the only strictly closed quantum system. 
Universality of quantum theory thus dictates that the Universe as a whole
has to be described by quantum theory -- this is the realm of
quantum cosmology. For its interpretation, no reference to
an external measurement agency can be made. Since such an interpretational
scheme provides insight into quantum theory in general, it was
claimed that ``quantum mechanics is best and most fundamentally understood
in the framework of quantum cosmology'' \cite{GMH}. 

The idea of quantum cosmology is more general than the quantization
of a particular interaction. However, since gravity dominates on large
scales, any reasonable formalism of quantum cosmology must employ
a quantum theory of gravity. 
Such a theory is not yet available in a definite form, but various
promising approaches exist \cite{OUP}. The main approaches are:
\begin{itemize}
\item {\em Superstring theory (M-theory)}: This is a unified quantum theory
of all interaction, from which quantum gravity emerges in an appropriate
limit.
\item {\em Quantum general relativity}: This is the application of
established quantization rules to general relativity. It may lead to 
a viable theory on the non-perturbative level or, at least, to
an effective theory away from the Planck scale. From a methodological point
of view, one can further subdivide this approach: one example is
the path-integral approach (Euclidean or Lorentzian), another example
is canonical quantum gravity. Depending on the chosen variables,
one can distinguish in the latter between, for example, quantum
geometrodynamics (`Wheeler--DeWitt equation') and loop quantum gravity.
\end{itemize}
Other, even more ambitious approaches, start with fundamental discrete
structures such as causal sets \cite{RS}. For the present discussion
it is sufficient to restrict to canonical quantum gravity, since it contains 
all conceptual tools that are required.  

The topic addressed here
is the observed irreversibility of the world and its possible justification
from quantum cosmology \cite{Zeh}.
I do not consider here the possibility of a new, fundamental, irreversible
law as discussed, for example, in \cite{NM} where a fundamental
master equation arises via the cosmological constant. Instead, 
I shall argue that the formal structure of the equations of canonical
quantumg gravity by themselves suggests a simple boundary condition
from where the arrows of time follow naturally. 
I shall start with a brief review of the quantum cosmological formalism
and the problem of the arrows of time. I then attempt to trace the
origin of these arrows to a simple boundary condition in quantum
cosmology. Finally I shall briefly discuss possible consequences for
black holes.

%%%%%%%%%%%%%%%%%%%%%%%%%%%%%%%%%%%%%%%%%%%%%%%%%%%%%%%%%%

\section{The formalism of quantum cosmology}

The central equation in canonical quantum gravity is the
quantum constraint equation \cite{footnote1}
\be
\lb{WDW}
\hat{H}\Psi=0\ ,
\ee
where $\hat{H}$ denotes the full Hamiltonian of gravitational and other
degrees of freedom. Equation (\ref{WDW}) is, mostly in the
geometrodynamical context, called the Wheeler--DeWitt equation.
Among the important properties of (\ref{WDW}) are:
\begin{itemize}
\item The quantum state $\Psi$ depends only on three-dimensional
quantities. It is invariant under three-dimensional 
coordinate transformations. 
\item No external (`non-dynamical') time parameter is present;
the state $\Psi$ describes a `stationary wave'.
\item An equation of the form (\ref{WDW}) follows from {\em any} theory
that is time-reparametrization invariant on the classical level.
\item In the geometrodynamical case, (\ref{WDW}) is pointwise
hyperbolic and thus defines an `intrinsic time'. In quantum
cosmological models, the intrinsic time is given by the
scale factor (or the spatial volume) of the Friedmann universe.
\end{itemize}
Consider the simple model of a (closed) Friedmann universe with
scale factor $a\equiv \exp(\alpha)$ and a homogeneous massive
scalar field $\phi$ with mass $m$. The corresponding Wheeler--DeWitt
equation then reads \cite{footnote2}
\be
\lb{Friedmann}
\left(G\hbar^2\frac{\partial^2}{\partial\alpha^2}
-\hbar^2\frac{\partial^2}{\partial\phi^2}+m^2\phi^2e^{6\alpha}
-\frac{e^{4\alpha}}{G}\right)\psi(\alpha,\phi)=0\ .
\ee
This equation should be valid at least for scales bigger than the
Planck scale: Since general relativity is the established classical
theory of gravity on large scales, and since we assume quantum theory
to be universally valid, the direct quantization of general relativity
should be valid at least as an effective theory on large scales.
It may, however, break down near the Planck scale. An {\em ad hoc}
modification can be made, for example, by the introduction of
an appropriate `Planck potential' in order to
facilitate the normalizability of the wave function
there \cite{CZ}.
A more fundamental approach would be to employ results from the
full theory before the restriction to homogeneous models. 
This can be achieved in loop quantum gravity where the spectra of
geometrical operators turn out to be discrete, cf. \cite{OUP,BMT}
and the references therein. 
For cosmological models one then finds that the Wheeler--DeWitt equation
(\ref{Friedmann}) is replaced by a {\em difference equation} for steps
characterized by $n\in{\mathbb Z}$ \cite{BMT}. The number $n$ is related
to the eigenvalue of the operator $\hat{p}$,
\be
\hat{p}\vert n\rangle=\frac{1}{6}\beta l_{\rm P}^2n\vert n \rangle \ ,
\ee
where $\beta$ denotes a quantization ambiguity (the `Barbero--Immirzi
parameter'), $l_{\rm P}=\sqrt{8\pi G\hbar}$ is the (reduced) Planck length,
and $\hat{p}$ is the operator corresponding to the classical quantity $p$,
where $\vert p\vert =a^2$. In the limit $n\gg 1$, the difference equation
goes over into the differential equation (\ref{Friedmann}). The number $n$
can be interpreted as playing the role of `discrete intrinsic time'. 

A most important question is how to address appropriate boundary conditions.
Since there is no external time, boundary conditions have to be imposed
with respect to intrinsic (dynamical) degrees of freedom.
For the Wheeler--DeWitt equation in quantum cosmology, the scale factor
presents itself as the appropriate timelike variable. Subtleties
arise for the case where the classical model describes a recollapsing
universe. If one wants to represent such classical solutions 
in the quantum theory by wave packets, the `returning packet' has to be
present `initially' (with respect to an `initial condition' for
$a=$ constant). Imposing this on solutions to (\ref{Friedmann}), 
it turns out that one {\em cannot} find a narrow wave packet all along
the classical trajectory -- the semiclassical approximation must breakdown
somewhere \cite{packets}. This breakdown is connected with the presence
of a turning point in the classical theory. 

What about boundary conditions in the case of loop quantum cosmology?
It turns out there that, for a particular factor ordering, the state
$\psi_0$ corresponding to $n=0$ in the difference equation drops out.
 The difference equation can thus be continued through the
`classical singularity' (which would be at $n=0$) into the regime of
negative $n$. Still, the equation that would contain $\psi_0$ has to be
fulfilled and leads to a constraint among the other $\psi_n$. 
This is interpreted as a `dynamical initial condition' \cite{Bo,BMT}.
The absence of the classical singularity is also recognized by the
fact that the inverse of the operator $\hat{a}=\sqrt{\vert\hat{p}\vert}$
is {\em bounded}.

Again, the case of a closed universe exhibits subtleties
because it seems that a divergent behaviour of quantum states
at large scale factor cannot be avoided \cite{GU}. The origin of this
problem is the fact that two possible orientations of the triads are
needed in the loop approach, providing the means to continue the
difference equation through $n=0$ and to avoid the classical singularity.

For the Wheeler--DeWitt equation, the question of singularity avoidance
can be rigorously discussed within simple models. One is the
case of a null dust shell that classically collapses to form a black hole. 
Demanding unitarity, a corresponding quantum theory can be constructed
that fully avoids the singularity -- collapsing wave packets enter
the Schwarzschild radius, but then re-expand to infinity, see 
\cite{OUP,PH} and the references therein. The full solution thus 
describes a superposition of the black-hole and white-hole situation
and is entirely singularity-free. 

An important issue for any approach to quantum gravity is the
semiclassical approximation. For the Wheeler--DeWitt equation, 
this can be achieved, at least on a formal level, by a
Born--Oppenheimer type of approximation scheme, with the non-gravitational
degrees of freedom adiabatically following the `slowly developing'
gravitational variables \cite{OUP}. While for full loop quantum gravity
this may not yet be clear, the situation in loop quantum cosmology is
straightforward: For $n\gg1$, the difference equation becomes identical
to the Wheeler--DeWitt equation, and the standard Born--Oppenheimer
scheme can then be applied. 

As is well known, one can recover from the Wheeler--DeWitt equation
in the semiclassical limit
a functional Schr\"odinger equation for the non-gravitational degrees
of freedom \cite{OUP}. The corresponding time parameter is defined
through the slowly evolving gravitational variables, typically
the expansion of the universe. An important ingredient is decoherence
of relevant variables (such as the volume of the universe) by
irrelevant variables (such as small density fluctuations). Otherwise
one would encounter superpositions of macroscopically different
universes. Decoherence `starts' with the onset of inflation; before 
inflation, the universe is timeless and there is no classical evolution
\cite{BKKM}.
In fact, due to the unavoidable quantum entanglement between
matter and gravity, mutual decoherence arises. This may mimic
a gravity-induced collapse of the wave function as discussed, for example,
in \cite{LD}. Because of decoherence, one must use a master equation instead
of the Schr\"odinger equation, valid for the evolution into the
forward direction of (semiclassical) time $t$. 
If this master equation holds in that direction of $t$ that corresponds
to increasing scale factor $a$, it cannot be valid across a
classical turning point into a recollapsing phase. This indicates
that the emergence of an arrow of time for a classically recollapsing universe
is subtle, see Section~IV.

%%%%%%%%%%%%%%%%%%%%%%%%%%%%%%%%%%%%%%%%%%%%%%%%%%%%%%%%%

\section{Arrows of time}

Most of the fundamental laws of Nature do not distinguish between past
and future, but there are many classes of phenomena that exhibit
an arrow of time \cite{Zeh}. This means that their time-reversed
version is, under ordinary conditions, never observed. 
Arthur Eddington called these classes `arrows of time'. The main arrows 
are the following:
\begin{itemize}
\item Radiation arrow (one observes retarded solutions of the wave
equation, but no advanced solutions);
\item Thermodynamical arrow (Second Law of thermodynamics, demanding
the entropy to be non-decreasing for a closed system);
\item Quantum mechanical arrow (measurement process and emergence
of classical properties by decoherence);
\item Gravitational arrow (expansion of the universe and emergence
of structure by gravitational condensation).
\end{itemize}
Since the expansion of the universe is a single process (not a class of
phenomena), it has been suggested (starting from the work of
Ludwig Botzmann) that it provides the root of all these arrows, the
`master arrow'. The various arguments that lead to this suggestion
are discussed in great detail in \cite{Zeh}. It is important in this
context to remind oneself that, because gravitational systems possess
negative heat capacity, homogeneous states are characterized by a {\em low}
gravitational entropy, whereas inhomogenous states have a {\em high}
gravitational entropy. This is just the opposite than for non-gravitational
systems. The maximal entropy would be reached for our universe if
all matter had collapsed into a single gigantic black hole. 
We are obviously far away from such a state. Our universe is thus
characterized by an extremely unprobable initial condition of low
gravitational entropy, or, in other words: Why did the universe
start so smoothly? 

One can try to answer this question within the classical theory.
Recent attempts include the suggestion that `eternal inflation' may
be responsible \cite{CN}. However, any fundamental attempt of explanation
should address this issue in the framework of quantum gravity which
transcends the classical theory. The immediate challenge to face is then
the `timelessness' of quantum gravity, cf. (\ref{WDW}). 
How can one derive an arrow of time from a framework that does
contain no time? I shall discuss in the following section how this
can be achieved, at least in principle.

%%%%%%%%%%%%%%%%%%%%%%%%%%%%%%%%%%%%%%%%%%%%%%%%%%%%%%%

\section{Origin of irreversibility from quantum cosmology}

Quantum gravity does not contain an external time parameter at
the most fundamental level. As discussed above, however, one can
introduce the concept of an intrinsic time, given in quantum cosmology
by the scale factor $a\equiv\exp(\alpha)$ 
of the universe (or its discretized version).
The important observation is now that the fundamental equation {\em is}
asymmetric with respect to $a$. Considering a Friedmann model
with small perturbations (symbolically denoted by $\{x_i\}$),
the Wheeler--DeWitt equation (\ref{WDW}) is of the form
\begin{equation}
\lb{WDW2}
\hat{H}\Psi=\left(\frac{\partial^2}{\partial\alpha^2}
+\sum_i\left[-\frac{\partial^2}{\partial x_i^2}+
V_i(\alpha,x_i)\right]\right)\Psi=0\ .
\end{equation}
The potential appearing in (\ref{WDW2}) is asymmetric with respect
to `intrinsic time' $\alpha$; one has, in particular, the important 
property that $V_i\to 0$ for $\alpha\to-\infty$. This allows 
one to impose a very simple boundary condition in this limit.
As Zeh has suggested \cite{Zeh}, one can demand that
\begin{equation}
\lb{bc}
\Psi\stackrel{\alpha\to -\infty}{\longrightarrow}
\chi(\alpha)\prod_i\psi_i(x_i)\ ,
\end{equation}
that is, an initial condition where the various degrees of freedom
are {\em not} entangled. Solving then the Wheeler--DeWitt equation
with this condition, entanglement automatically {\em increases}
with increasing $\alpha$. This, then, leads to an increase of
the entropy for the relevant degrees of freedom which include
the scale factor and some additional relevant variables
$\{y_i\}$. The entropy is found from 
\be
S(\alpha,\{y_i\})=-k_{\rm B}{\rm tr}(\rho\ln\rho)\ ,
\ee
where $\rho$ is the reduced density matrix obtained by integrating out
all irrelevant degrees of freedom from the full quantum state. 
This increase of entropy then {\em defines} the direction of time. 
All the arrows of time discussed in Section~III would then have
their common root in this entropy increase. 
The emergence of a correlated state from the symmetric initial state
(\ref{bc}) then represents a `spontaneous symmetry breaking'
similar to the symmetry breaking when the quantum field theoretic
vacuum proceeds from a symmetric to an asymmetric state
\cite{Zeh}, cf. also \cite{CZ}.
Since the time parameter $t$ in the semiclassical limit is defined
as a function of the scale factor, time is defined by the expansion of
the universe. In a sense, the expansion of the universe is a
tautology. It would be extremely interesting to perform the
above analysis for the difference equation of loop quantum cosmology.  

What happens for a universe that is classically recollapsing?
Since the boundary condition (\ref{bc}) is formulated for
$\alpha\to-\infty$, irrespective of any classical trajectory,
it applies at the same time to the `big bang' and the `big crunch'. 
Only one condition is thus needed in order to cover both regions.
Consequently, increase of entropy is always correlated with 
increase of scale factor, that is, increasing size of the universe.
But what happens at the turning point? There the arrow of time
reverses, but the reversal is only of formal significance. 
Since the semiclassical approximation breaks down there \cite{packets},
the universe is fully quantum in this region -- no classical
observer could survive it \cite{KZ}. Many quasi-classical components
of the full quantum state, each representing a universe of its own,
interfere there destructively in order to fulfill the final condition
of the wave function going to zero for $\alpha\to\infty$. 
Quantum cosmology thus not only specifies the beginning of the
classical evolution (when decoherence sets in at the onset of inflation),
but also its end.

%%%%%%%%%%%%%%%%%%%%%%%%%%%%%%%%%%%%%%%%%%%%%%%%%%%

\section{Consequences for black holes}

A fundamental quantum cosmological boundary condition such as
(\ref{bc}) has also profound consequences for black holes
in a recollapsing universe
\cite{Zeh,KZ}. Consider an object, for example a dust shell or a star,
that collapses to form a black hole (assume a Schwarzschild black hole,
for simplicity). The collapse is supposed to happen (in the proper
time of the collapsing object) long before the universe as a whole 
reaches its maximum expansion at, say, a Schwarzschild time
$t_{\rm turn}$. Since the full quantum cosmological boundary condition
is symmetric, the collapsing object must expand again for
$t>t_{\rm turn}$, although any observer would experience this as collapsing
(because his arrow of time always points from small $a$ to large $a$).
Because of the mentioned quantum effects near the turning point,
no classical connection exists between the collapse of the object and
its following (formal) expansion. As a consequence, one has \cite{KZ}
\begin{itemize}
\item no horizon formation,
\item no singularity inside the black hole, and therefore
\item no information-loss problem and
\item no need to introduce cosmic censorship, since
 also no naked singularities form.
\end{itemize}
Unfortunately, these consequences cannot be tested from outside
(due to the large redshift), but only by volunteers plunging into the
black hole -- they would enter the quantum era of the cosmological
turn-around within a short proper time. 

Recently, an {\em ad hoc} final state boundary condition 
was imposed at black-hole singularities in order to prevent
information from being absorbed by the singularity \cite{HM}.
Like in the case discussed here, the corresponding quantum state
consists of a superposition of many macroscopically distinct states.
However, in our case this consequence follows directly from 
the fundamental framework of quantum cosmology -- and the 
`information-loss problem' does not exist because neither a horizon nor
a singularity would ever form. A final answer can, of course, only
be obtained after the full theory of quantum gravity has been
constructed and experimentally tested.

%%%%%%%%%%%%%%%%%%%%%%%%%

\section{Acknowledgments}

I would like to thank the organizers
of DICE2004, and in particular Hans-Thomas Elze,
for inviting me to this wonderful and inspiring meeting.

%%%%%%%%%%%%%%%%%%%%%%%

\bibliography{apssamp}% Produces the bibliography via BibTeX.

\begin{thebibliography}{00}

\bibitem{deco} E. Joos, H. D. Zeh, C. Kiefer, D. Giulini, J. Kupsch,
and I.-O. Stamatescu, {\em Decoherence and the Appearance of a Classical
World in Quantum Theory}, second edition (Springer, Berlin, 2003).
See also {\tt www.decoherence.de}.

\bibitem{GMH} M. Gell-Mann and J. B. Hartle, in: {\em Complexity, Entropy,
and the Physics of Information}, edited by W. H. Zurek
(Addison-Wesley, Reading, 1990). 

\bibitem{OUP} C. Kiefer, {\em Quantum Gravity} (Clarendon Press,
Oxford, 2004).

\bibitem{RS} R. Sorkin, {\tt gr-qc/0309009}, see also his
contribution to these Proceedings.

\bibitem{Zeh} H. D. Zeh, {\em The physical basis of the direction of
 time}, fourth edition (Springer, Berlin, 2001). See also
{\tt www.time-direction.de}.

\bibitem{NM} N. E. Mavromatos, {\tt gr-qc/0411067}, contribution to these
Proceedings.

\bibitem{footnote1}  This is the form for a closed
universe. In the open case (relevant e.g. for black holes), boundary
terms are present \cite{OUP}.

\bibitem{footnote2} Units are chosen such that $c=1$ and
$G\to (3\pi/2)G$.

\bibitem{CZ} H. D. Conradi and H. D. Zeh, {\em Phys. Lett. A} {\bf 154},
321 (1991).

\bibitem{BMT} M. Bojowald and H. A. Morales-T\'ecotl, {\tt gr-qc/0306008}.

\bibitem{packets} C. Kiefer, {\em Phys. Rev. D} {\bf 38}, 1761 (1988).

\bibitem{Bo} M. Bojowald, {\em Phys. Rev. Lett.} {\bf 87}, 121301 (2001).

\bibitem{GU} D. Green and W. G. Unruh, {\tt gr-qc/0408074}.

\bibitem{PH} P. H\'aj\'{\i}\v{c}ek, in: {\em Quantum gravity:
from theory to experimental search}, edited by D. Giulini, C. Kiefer, and
C. L\"ammerzahl (Springer, Berlin, 2003).

\bibitem{BKKM} A. O. Barvinsky, A. O. Kamenshchik, C. Kiefer, and
I. V. Mishakov, {\em Nucl. Phys. B} {\bf 551}, 374 (1999).

\bibitem{LD} L. Di\'osi, {\tt quant-ph/0412154},
 contribution to these Proceedings.

\bibitem{CN} S. M. Carroll and J. Chen, {\tt hep-th/0410270};
H. Nikoli\'c, {\tt hep-th/0411115}.

\bibitem{KZ} C. Kiefer and H. D. Zeh, {\em Phys. Rev. D} {\bf 51}, 4145 (1995).

\bibitem{HM} G. T. Horowitz and J. Maldacena, {\em JHEP} 0402 (2004) 008.

\end{thebibliography}
%%%%%%%%%%%%%%%%%%%%%%%

\end{document}